\documentclass[RNAAS]{aastex631}

\graphicspath{{./}{figures/}}

\begin{document}

\title{On the Chromaticity of the (NEO)WISE Astrometry}

\correspondingauthor{Bringfried Stecklum}
\email{stecklum@tls-tautenburg.de}

\author[0000-0001-6091-163X]{Bringfried Stecklum}
\affil{Thüringer Landessternwarte Tautenburg,
Sternwarte 5,
Tautenburg, 07778, Germany}

\begin{abstract}
The Wide-field Infrared Survey Explorer (WISE) \citep{2010AJ....140.1868W} and its follow-up Near-Earth Object (NEO) mission (NEOWISE, \cite{2011ApJ...743..156M}) scan the mid-infrared sky twice a year. The spatial and temporal coverage of the resulting database is of utmost importance for variability studies, in particular of young stellar objects (YSOs)
which have red $W1{-}W2$ colors.
During such an effort, I noticed subarcsecond position offsets between subsequent visits. The offsets do not appear for targets with small $W1{-}W2$ colors, which points to a chromatic origin in the optics, caused by the spacecraft pointing alternating ``forward'' and ``backward'' from one visit to another. It amounts to 0\farcs1 for targets with $W1{-}W2\approx2$.
Consideration of this chromatic offset will improve astrometry.
This is of particular importance for NEOs that are generally red.
\end{abstract}


\section{Introduction} \label{sec:intro}

The cryogenic WISE space telescope was launched in 2009 into a Sun-synchronous orbit. It features a 40-cm mirror for all-sky imaging at mid-infrared (MIR) wavelengths. 
After coolant run-out, the satellite was hibernated in 2011 and reactivated in 2013 for the NEOWISE mission. Its primary focus is on the detection of NEOs. Due to its orbit, it visits a sky region twice a year. 
Despite the sparse temporal sampling, the temporal coverage of more than ten years promotes the (NEO)WISE data to be a treasure trove for YSOs variability studies.
Due to their spectral energy distribution (SED), YSOs have a red color index for the W1 (3.4\,\micron{}) and W2 (4.6\,\micron) (NEO)WISE bands.

(NEO)WISE data are of particular importance for identifying and monitoring episodic accretion events of YSOs that allow us to gain insight into short and intense episodes of protostellar growth. In this regard, numerous studies have been conducted, e.g., \cite{2013MNRAS.430.2910S}, \cite{2019MNRAS.483.4424K}, \cite{Stecklum:2017}, and \cite{2020MNRAS.499.1805L}. Such events can also be periodic, most likely due to modulation of the accretion flow in a protobinary system. 
This scenario might hold for several objects, e.g., \cite{Stecklum:2018} and \cite{Kobak:2023}.

\section{Data retrieval and analysis}
During recent years, (NEO)WISE data for YSOs have been gathered by the author from the NASA/IPAC Infrared Science Archive (IRSA)
to study their variability. Usually, (NEO)WISE photometry and positions were retrieved within 5\arcsec{} of the YSO locations, based on frames with the photometric quality flag A. 
Because the WISE instrumentation reached a new thermal equilibrium after the coolant was consumed, the present analysis is based only on the NEOWISE positions.  

\begin{figure}   
\centering
\includegraphics[width=.95\textwidth]{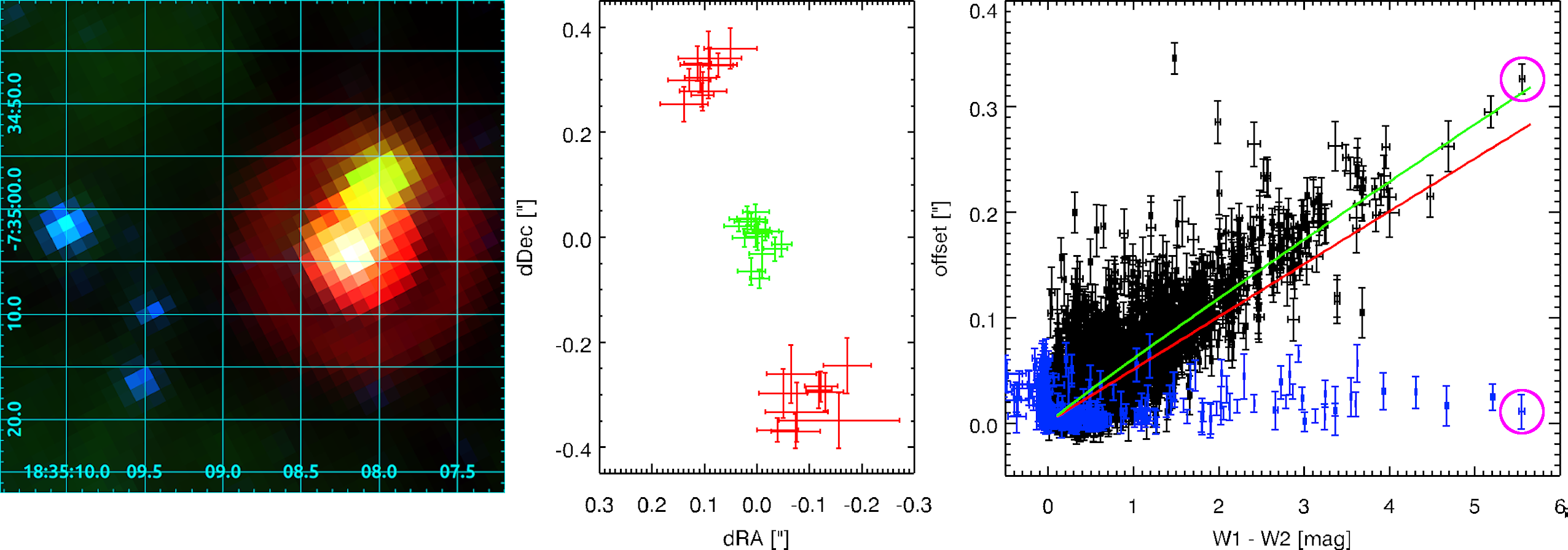}
\caption{{\bf Left:} MIR RGB image covering \object[{[HMW2016] G024.328+00.14}]{IRAS 18324-0737} (right border) and the comparison star 2MASS~18350998-0735015 (left border).
{\bf Center:} Red - mean positions for the YSO are symmetrically displaced along the scan direction for forward- and backward-looking visits, green - same for star 29\arcsec{} east of the YSO. The chain-like distribution of the YSO positions is due to variability that affects its photocenter location \citep{Kobak:2023}. {\bf Right:} $W1{-}W2$ dependence of the offset and linear (red) as well as power law (green) approximations. Blue symbols denote positions for which offset correction was applied. The purple circles mark \object[{[HMW2016] G024.328+00.14}]{IRAS 18324-0737} before and after correction.}
\label{fig:wchroma}
\end{figure}

Among the objects studied is the massive YSO \object[{[HMW2016] G024.328+00.14}]{IRAS 18324-0737} that seems to show periodic brightness bursts. With $W1{-}W2=5.55\pm0.42$ it is the reddest YSO of the sample. Its mid-infrared (MIR) appearance is shown in the left panel of Fig.\,\ref{fig:wchroma} which is based on MIPS 24\,\micron{} (red), IRAC 8\,\micron{} (green) and 4.5\,\micron{} (blue) images. The yellowish feature arises from molecular emission. During the investigation, it became obvious that its NEOWISE coordinates form two clusters, offset to either side of the nominal position. This is shown in the central panel of Fig.\,\ref{fig:wchroma} where the mean positions and their errors per visit are plotted. The red symbols refer to \object[{[HMW2016] G024.328+00.14}]{IRAS 18324-0737} while the green symbols refer to a nearby reference source,  2MASS~18350998-0735015 ($W1{-}W2={-}0.06\pm0.02$), which appears blueish on the left border of the left panel. For both, the nominal position is at [0,0]. The absence of coordinate offsets for this source indicates that their origin is probably due to a chromatic effect of the WISE optics; cf. \cite{2005SPIE.5904..178S}. 

Aberrations in the optical path of an astronomical camera imply that the position of the stellar images is wavelength dependent. Thus, astrometry has to be corrected for this chromatic aberration through knowledge of the SED of the observed objects, e.g., \cite{2006MNRAS.367..290J}.  Dipole residuals suggestive of astrometric misalignments have already been observed for (NEO)WISE. But they were attributed to PSF models that are not symmetric with respect to swapping the scan direction \citep{2018AJ....156...69M}. A possible chromatic origin was considered, but was not explored in depth.

To derive the color dependence of the offset, 7333 mean visit positions with offset accuracies better than 0\farcs025 from before mid-2021 have been used. The right panel in Fig.\,\ref{fig:wchroma} shows offsets $o$ vs. $W1{-}W2$ color $\Delta W_{12}$. Black symbols refer to uncorrected values, while blue symbols represent corrected values. These comprise 369 mean visit positions, obtained after having implemented a correction method.

\section{Results and conclusions}
 The offset-color dependency is approximately linear (red line) but deviates for larger colors. A power law ${o}=a{\cdot}\Delta W_{12}^b$ (green line) is a better representation across the entire color range with the following parameters $a=0.061\pm0.011, b=0.954\pm0.006$. For the majority of NEOs in \cite{2011ApJ...743..156M} having $W1{-}W2$ of 1$\dots$3\,mag, this implies offsets in the order of 0\farcs06$\dots$0\farcs17.

When assessing the importance of this effect for NEOs, it must be kept in mind that NEOWISE observations during a visit lasting about a day usually yield a relatively short tracklet. It will be shifted with respect to the correct coordinates as a whole according to the $W1{-}W2$ color of the object. 
This systematic deviation might become obvious when ground-based measurements are incorporated to cover a larger arc of the orbit. Then the orbital solution will probably show residuals of the NEOWISE tracklet that exceed the nominal position accuracy. Since it is of utmost importance to estimate NEO orbits as accurately as possible, correction of the chromatic offset is recommended before submitting the position measurements.

Regarding YSOs, this correction allows for the detection of photocenter shifts that indicate photometric variability. Apart from the example of \object[{[HMW2016] G024.328+00.14}]{IRAS 18324-0737} mentioned above, an increasing number of objects showing similar shifts have been identified in the (NEO)WISE data. YSOs in early evolutionary stages are still surrounded by envelopes that feature bipolar outflow cavities. Alternating or erratic photocenter shifts, aligned with the position angle of the cavities, result from their uneven variable illumination. For geometrical reasons, this effect is most pronounced when they are oriented close to the plane of the sky. Its observed time frame in the range of years or even less indicates an origin of the modulation of radiation in the inner circumstellar disk.
Thus, (NEO)WISE astrometry enables us to improve our knowledge on protostellar variability.

In general, the odd-even technique for imaging in Sun-synchronous orbits can be applied to derive astrometric offsets due to optical aberrations.

\begin{acknowledgments}
This research has made use of the NASA/IPAC Infrared Science Archive, which is funded by the National Aeronautics and Space Administration and operated by the California Institute of Technology.

Thanks to Aaron Meisner for clarifications on his 2018 paper.
\end{acknowledgments}

\facilities{ NEOWISE, Spitzer}
\software{Data analysis was done using Interactive Data Language (IDL) (Exelis Visual Information Solutions, Boulder, Colorado).}

\bibliographystyle{aasjournal-hyperref}
\bibliography{wchroma}
\end{document}